\documentstyle[preprint,aps]{revtex}

\begin{document} 
\draft
\title{Quadrupole-quadrupole interaction calculations which include
N=2 mixing}
 
\author{M.S. Fayache,$^{1,3}$ L. Zamick,$^{2,3}$ and
Y.Y. Sharon$^3$\cite{byline}}
\address{(1) D\'{e}partement de Physique, Facult\'{e} des Sciences de
Tunis, Tunis 1060, Tunisia\\
\noindent (2) TRIUMF, 4004 Wesbrook Mall, Vancouver, B.C., CANADA V6T
2A3\\ 
\noindent (3) Department of Physics, Rutgers University, New
Brunswick, NJ 08903 USA}

\date{\today}
\maketitle

\begin{abstract}
We carry out a study of the study of the $\rm Q \cdot Q$ interaction
in a model space which consists of several nucleons in an open shell
and all $\rm 2\hbar \omega$ excitations. This interaction is $\rm -t
\frac{X_o}{2} Q \cdot Q$, where for t=1 we get the `accepted
strength'.
In the $0p$ space, the spectrum would scale with $t$. In this
space, the $2^+_1$ and $2^+_2$ states  of $^{10}$Be are degenerate, as
are the [330] 
and [411] sets of $J=0^+,~1^+$ and $2^+$ triplets. When $2 \hbar
\omega$ admixtures are included, the degeneracies are removed. For $t 
\geq 1.8$ we have new ground state and a
new $2_1^+$ state. These are states in which two particles are excited 
from the $0p$ to the $1s-0d$ shell. 
There is no mixing of these 2p-2h states with the other states. For
these 2p-2h states the occupancy for 0s,0p,1s-0d and 1p-of are 4,4,2
and 0 respectively.
\end{abstract}

\narrowtext

\section{Introduction}

In this work we will study the behaviour of the $\rm Q \cdot Q$
interaction in a nucleus as a function of the strength of the
interaction. Although this is a model study it helps to make things
concrete by focusing on a particular nucleus. We choose $^{10}Be$.
This nucleus is of 
particular interest because in a $0p$ space calculation with $\rm Q \cdot
Q$ there are some interesting degeneracies. For example, the $2_1^+$
and $2_2^+$ states are degenerate, and both have orbital symmetry
[400]. The $L=1~S=1$ states with orbital symmetry [330] and [411] are
degenerate. Thus we have two sets of degenerate triplets $J=0^+,~1^+$
and $2^+$ emanating from the above two orbital configurations. These
degeneracies can easily be found by applying the $SU(3)$ formula:

\[E(\lambda \mu)=\bar{\chi}\left[-4(\lambda^2+\mu^2+\lambda\mu+ 3
(\lambda+\mu))+3L(L+1) \right] \]

If we write the interaction as $\chi Q \cdot Q$ then, in the $0p$
space and when we change $\chi$, the degeneracies will not be
removed. All that will happen is that the energies of all states will
be proportional to $\chi$, and so the spectrum will be blown up or
shrunk as $\chi$ is made larger or smaller. 

What happens though if we include contributions from other major
shells? The $\rm Q \cdot Q$ interaction will not connect $\Delta N=1$
states because of parity arguments, but there will be $\Delta N=2$
admixtures. Indeed, the concept of an $E2$ effective charge is often
illustrated by using a $\rm Q \cdot Q$ interaction to obtain $\Delta N=2$
admixtures. 

\section{The Hamiltonian and Choice of Parameters}

It has been shown by Bes and Sorensen~\cite{bes} that if in the 
valence space ($0p$) the appropriate $\rm Q \cdot Q$ interaction is
$-\chi_0 Q \cdot Q$, then in a space which includes $\Delta N = 2$
excitations the appropriate strength is $-\frac{\chi}{2} Q \cdot
Q$. We wish to vary the interaction, and we therefore parameterize it
as 

\[V=-t \left (\frac{\chi_0}{2} \right ) Q \cdot Q \]

\noindent so that for $t=1$ we have the standard choice of Bes and
Sorensen~\cite{bes}. For $^{10}Be$, we have $\chi_0=
0.36146~MeVfm^{-1}$.  

The Hamiltonian we use is 

\[ H=\sum_i T_i + \frac{1}{2}m \omega^2 r_i^2 - \sum_{i < j} t
\frac{\chi_0}{2} Q(i) \cdot Q(j) \]

\noindent We perform shell model calculations using the OXBASH shell
model code~\cite{oxbash} in the space $(0p)^6$ plus $2 \hbar \omega$
excitations. As mentioned in the introduction, when harmonic
oscillator wave functions are used in a single major shell, the single
particle terms in the above Hamiltonian are constant, so 
that the only part of the Hamiltonian that affects the spectra is the
two-body term. The separation of energies is linear in $t$ -the wave
functions are unaffected by the choice of (positive) $t$. 

However, when $\Delta N =2$ excitations are allowed, the linear terms
are no longer constant, and the behaviour as a function of $t$ is more
complicated. We will now study this behaviour as a function of $t$ in
$^{10}Be$. 

\subsection{Removal of Degeneracies}

In Table I we present $t=1$ results for $^{10}Be$ in a large space
$(0p)^6$ plus all $2 \hbar \omega$ excitations. We present the results
for the energies of $J=1^+$ and $2^+$ $T=1$ states, as well as $B(M1)$
and $B(E2)$ transitions from the ground state to these states. More
precisely, it is the isovector $B(M1)$ in units of $\mu_N^2$ and
isoscalar $B(E2)$ ($e_p=1,~e_n=1$) in units of $e^2 fm^4$. 

We see that there are many degeneracies still present in the large
space $e.g.$ four $J=1^+~T=1$ states at 12.12 $MeV$ and three at 13.90
$MeV$, as well as three $J=2^+~T=1$ states at 12.12 $MeV$. These
degeneracies clearly correspond to various $S,~T$ combinations for
states of given $L$ and orbital symmetry [$f$]. 

However, other `accidental' degeneracies which were present in the
small space $(0p)^6$ are no longer present. The $J=1^+_1$ and $1^+_2$
states at 3.74 $MeV$ and 7.31 $MeV$ are linear combinations of the
$L=1$ [330] and $L=1$ [411] configurations (actually they are parts of
$J=0^+,~1^+,~2^+$ triplets). In the small space, these two $J=1^+$
states (or triplets) were degenerate -now one of the states is almost
at twice the excitation energy of the other.

The $2_1^+$ and $2_2^+$ states are no longer degenerate. The $2_1^+$
state is at 2.19 $MeV$ with $B(E2)\uparrow$ from the ground state of
63.8 $e^2 fm^4$, whilst the $2_2^+$ state is at 3.40 $MeV$ with
$B(E2)\uparrow=113.4~e^2 fm^4$. Note that, contrary to experiment, the
second $2^+$ state is the one most strongly excited. When a reasonable
spin-orbit interaction is added to the Hamiltonian the situation is
corrected. 

It should be pointed out that in perturbation theory, in which only
the {\em direct} part of the particle-hole interaction of $\rm Q \cdot Q$
is used to renormalize the interaction between two particles in the 
valence space, the degeneracies above would {\em not} be removed. The
relevant diagram is the familiar Bertsch-Kuo-Brown bubble (or phonon)
exchange between two nucleons~\cite{bert,kbrown}. For a simple $\rm Q
\cdot Q$ interaction, this diagram simply renormalizes the strength of
the $\rm Q \cdot Q$ interaction. Clearly, changing the strength in the
valence space will not remove the degeneracies. 
Thus, the shell model diagonalization implicitly contains effects
beyond the direct bubble diagram. Furthermore, these effects are quite
important. 

\subsection{Change in the Nature of the Ground State as $t$ Increases}

We now vary $t$ over the range $0 < t < 2$. In Fig. 1 we plot as a
dot-dash curve the value of $E/t$ for the lowest $2^+$ state, the one with
finite but small $B(E2)$ strength from ground. We also plot $E/t$ as a
solid line for the state with the strongest $B(E2)$ from ground. It
starts off at $t=1$ as the second $2^+$ state. In the $0p$ space, the
$2_1^+$ and $2_2^+$ states would be degenerate, and the curve for
$E/t$ vs. $t$ would be a horizontal line. However, in the
$0p+2\hbar\omega$ space there is a dependence on $t$ (and more so for
the solid curve). 

But, for $t\approx 1.8$ and beyond, all the $B(E2)$ strength from
ground state goes to the new lowest $2^+$ state. Furthermore, the
value of $E/t$ becomes constant for $t \geq 1.8$ $i.e.$ the curve
becomes horizontal. Clearly, the nature of the ground state changed
beyond $t=1.7$. We will now examine this change in more detail. 

If the only thing that happened was that there was a new $J=0^+$
ground state, then of course there would be a sudden change in the
$B(E2)\uparrow$'s from this new ground state to the $2^+$
states. However, the static quadrupole moments of the $2^+$ states
themselves would not change. 

For $t=1.1$ the $B(E2)\uparrow$ to the $2_1^+$ state is 40.61 $e^2
fm^4$ and to the $2^+_2$ state 158.0 $e^2 fm^4$. The calculated
static quadrupole moments are respectively 11.56 $e fm^2$ and -11.46
$e fm^2$. As discussed by Fayache, Sharma and Zamick in
Ref.~\cite{qqt}, this is consistent with the two $2^+$ states being
prolate, with about the same intrinsic quadrupole moment $\rm Q_0$, but
the lower state would have $K=2$ and the upper one $K=0$. Indeed, for
$K=0$ $\rm Q(2^+)=-2/7Q_0$, and for $K=2$ $\rm Q(2^+)=+2/7Q_0$. 

The behaviour for $t=1.1$ is maintained for $t=1.3$ and 1.5, but for
$t=1.7$ and up to $t=1.75$ there is a big drop in $B(E2)_{0_1^+
\rightarrow 2^+_2}$. The other three quantities do not change, not even
-strangely enough- $\rm Q(2^+_2)$.

Remember that for $t=1.75$ we are still below the critical $t$ for
which the $J=0^+$ ground state changes its nature. What is clearly
happening is that a third $2^+$ state has crossed over and came below
what was formerly the $2^+_2$ state. This is confirmed by noting that
at $t=1.75$ the $B(E2)_{0^+_1 \rightarrow 2^+_3}$ is very large (162.1
$e^2 fm^4$). Clearly, in going from $t=1.7$ to $t=1.75$, the $2^+_2$
and $2^+_3$ states have interchanged positions. The fact that the
static quadrupole moments of the two states are about the same means
that both states can be associated with two different prolate $K=0$
bands which have the same deformation.

Next we consider $t=1.76$. Here we are just beyond the critical $t$,
and the nature of the ground state has changed. Now the $B(E2)$ to the
$2^+_1$ state is much weaker, and the $B(E2)$ to the $2^+_2$ state is
strong. This suggests that the new $0^+$ ground state and the new
$2^+_2$ state are members of a new rotational band, and that both of
these states have come down in energy together.

When we go from $t=1.76$ to $t=1.78$ there is a big change, but the
results stabilize beyond that. Now the $B(E2)$ to the $2_1^+$ state is
the strongest (140 $e^2 fm^4$), and the {\em signs} of the static
quadrupole moments change. Now $\rm Q(2^+_1)$ is negative, and $\rm Q(2^+_2)$
is positive. 

What is clearly happening is that there is another cross-over. What
was formerly the $2^+_3$ state at $t=1.7$ first crosses the the
$2^+_2$ state at $t=1.75$ (as mentioned above), and now crosses the
$2^+_1$ state at $t=1.78$. By $t=1.78$ and beyond, the $0^+$ and $2^+$
members of a new band have become the lowest two states, and the
results stabilize. 

\section{Interpretation of the New Band: States with Integer Occupancies}

In Fig. 2 we plot the rapid descent of the $J=0^+$ and $2^+$ members
of the new rotational band. We start from $t=1$, but if we project
backward we see that for small $t$ the band emanates from the $2 \hbar
\omega$ region. To better ascertain the nature of the new band, we 
give in Table III the occupancies of the single-particle levels that
were used in this calculation {\em i.e.} $0s$, $0p$, $1s-0d$ and
$1p-0f$. 

At $t=1.7$, just before the critical value, the $J=0^+$ ground state
is normal. The occupancy of the four major shells (in the order
mentioned above) is 3.84, 5.78, 0.18 and 0.20. The first excited $0^+$
state at 0.585 $MeV$ has occupancy 4, 4, 2 and 0. Clearly two nucleons
have been excited from $0p$ to $1s-0d$. 

What is at first surprising is that, for this state, the occupancies
are {\em precisely integers}. This is not an isolated example. It is
also true at $t=1.7$ for the second $2^+$ state at 4.33 $MeV$. 

When we  go to $t=1.9$, we have passed the critical value, and things
have settled down. the lowest $0^+$ and $2^+$ states are now the
$2p-2h$ states, both with the integer occupancies 4, 4, 2 and 0. 

Whereas most states do not have integer occupancies, there are many
which do. These are at higher energies. For example, for $t=1.7$,
there are other states with the occupancy 4, 4, 2, 0 at 21.64, 22.95,
22.97, 25.32 and 35.75 $MeV$. These are states with occupancy (3, 4,
1, 0) at 43.6 and 44.2 $MeV$. The latter correspond to lifting one
nucleon through two major shells. 

Why do we get such a simple behaviour for the $2p-2h$ states? The
answer involves a special feature of the $\rm Q \cdot Q$ interaction: all
matrix elements in which two particles in a major shell $N$ scatter
into a major shell $N \pm 1$ vanish. This is due to a parity selection
rule. For example, $\langle 0p~0p | Q\cdot Q | 0d~0d \rangle$ factors
into $\langle 0p |Q | 0d \rangle \langle 0p |Q | 0d \rangle$, and each
of these factors vanishes because of this parity rule.

Carrying the argument further, there can be no matrix element coupling
the $(0s)^4(0p)^4(0d-1s)^2$ configuration with other configurations
such as $(0s)^4(0p)^6$ or $(0s)^4(0p)^5(0f-1p)$ etc... 

Also, in our calculation we have limited the space to 2 $\hbar \omega$
excitations. Once we create the state with occupancy (4,4,2,0) our model
space does not permit further excitations. This explains the integer
occupancy. Presumably if we enlarged the space to include 4 $\hbar
\omega$ excitations we would no longer have the integer occupancies. It
would also be of interest to study the $4p-4h$ states.

This also explains why, as we increase $t$ towards 1.76, the descent
of the new band is so simple. Since there is no mixing with the other
configurations, the $2p-2h$ $J=0^+$ and $2^+$ states can just slip
down below the $(0p)^6$ states.

The rapid descent of the $2p-2h$ states can be understood in terms of
the Nilsson model. To form the $2p-2h$ state, we take two nucleons
from the $0p$ shell and put them in the Nilsson orbit ($Nm_3\Lambda$)
with quantum numbers (220). This orbit comes down rapidly in energy as
the nuclear deformation is increased. The Nilsson one-body deformed
Hamiltonian can be obtained from the $\rm Q \cdot Q$ two-body interaction
by replacing $\rm Q \cdot Q$ by $\rm Q \cdot \langle Q \rangle$ where $\langle
Q \rangle$ is the quadrupole moment of the intrinsic state. 

\section{Closing Remarks}

In this work, we have studied the properties of the interaction 
$-t\frac{\chi_0}{2} Q \cdot Q$ as a function of the coupling strength
$t$ in an extended model space which includes all $2 \hbar \omega$
excitations beyond the valence space. Using $^{10}Be$ as an example,
we found that states that were `accidentally' degenerate in the $0p$
valence space ($e.g.$ $2_1^+$ and $2_2^+$ of orbital symmetry [42], or
the $J=0^+,1^+,2^+$ triplets of orbital symmetries [411] and [33]),
are no longer degenerate in the extended space. This means that the $2
\hbar \omega$ admixtures do more than renormalize the coupling
strength of $\rm Q \cdot Q$.

The extended model space allows for $2p-2h$ admixtures and indeed, for
sufficiently large $t$ ($t \geq 1.8$), the $J=0^+$ and $2^+$ members
of this new band become the new $0_1^+$ and $2_1^+$ states. We find 
that this band, unlike the `normal' ground state band for $t < 1.7$,
has integer occupancies 4, 4, 2 and 0 for $0s$, $0p$, $1s-0d$ and
$1f-0p$ respectively. There is no mixing between the new $2p-2h$ band
and the $0p-0h$ band. This is a special feature of the $\rm Q \cdot Q$
interaction. 

On a speculative level we may argue that, for the $2p-2h$ band 
above, we should use a value of $t$ considerably larger than 1. We
don't have enough model space to renormalize the two-body interaction
in this band. Just as the interaction between the $(0p)^6$ states
gets renormalized by the configurations in which one nucleon is
excited through 2 major shells, so the interaction for the $2p-2h$
state would get renormalized by allowing at least one nucleon to be
excited through 2 major shells. But this would be a 4 $\hbar \omega$
state which, for practical reasons, we don't have in our model space.

We can use the $\rm Q \cdot Q$ interaction to place these $2p-2h$ bands at
the correct energies, but we will need other components of the
realistic nucleon-nucleon interaction in order to mix these bands with
the $0p-0h$ bands. For example, we could use the dipole-dipole or
octupole-octupole parts of the nucleon-nucleon
interaction. Alternatively one can work directly with realistic
interactions. 

There has been much progress in large-basis shell model calculations
in light nuclei. For example, there is the work of W.C. Haxton and
C. Johnson~\cite{hax} where they actually get the superdeformed
$4p-4h$ state in $^{16}O$ at a reasonable energy, although perhaps not
with the full quadrupole collectivity. There is also the work of Zheng
$et. al.$~\cite{zheng} where up to $8 \hbar \omega$ excitations have
been included in calculations of nuclei ranging from $^4He$ to
$^7Li$. Also, Zamick, Zheng and Fayache~\cite{zzf} required
multi-shell admixtures to demonstrate the `self-weakening mechanism'
of the tensor interaction in nuclei.

Nevertheless, schematic interactions like $\rm Q \cdot Q$ still play a
primary role in describing nuclear collectivity throughout the
periodic table. They are of special importance for highly deformed
intruder states. 

Surprisingly, there have been very few studies of schematic
interactions in multi-shell spaces. In the Elliott $SU(3)$ model,
momentum terms have been introduced to prevent $N=2$ admixtures in the
valence space~\cite{elliott}. This has lead to great simplicities and
beautiful results. There have been $R.P.A.$ studies with $\rm Q \cdot Q$
which involve $N=2$ mixing. These studies give $E2$ effective charge
renormalizations in the valence space and also the energies of the
giant quadrupole resonances, but they will not give us the highly
deformed states such as the $2p-2h$ state that we have found here. 
We therefore feel that careful studies of the schematic interactions
in multi-shell spaces are important, and we hope that others will
agree. 

\acknowledgements

This work was supported by a Department of Energy grant
DE-FG02-95ER 40940. One of us (L.Z.) thanks E. Vogt, B. Jennings, 
H. Fearing, and P.Jackson for their help and hospitality at TRIUMF.

\begin{figure}
\caption{The ratio $E(2^+)/t$ for the $2^+$ state with the strongest
$B(E2)$ (solid line, usually the second $2^+$ state), and the same
ratio for the lowest $2^+$ state (dash-dot line). For $t > 1.77$ all the
strength goes into one state (solid line).}
\end{figure}

\begin{figure}
\caption{The energies of the $0^+$ and $2^+$ members of a $2p-2h$
band, which descend rapidly as $t$ is increased. This band becomes the
ground state band beyond $t=1.78$, and in this model space has integer
occupancy (4,4,2,0).}
\end{figure}

\begin{table}
\caption{Energies of States in $^{10}Be$ in a Large Space Calculation
and Transitions from the Ground State}
\begin{tabular}{cc}
$E(J=1^+~T=1)~(MeV)$ & $B(M1)\uparrow$\tablenotemark[1] $(\mu_N^2)$\\ 
\tableline
 3.74  & 0.0\\
 7.31  & 0.0\\
 12.12 & 0.071\tablenotemark[2]\\
 13.90\tablenotemark[3] & 0.0\\
 22.10 & 0.0\\
\tableline
\tableline
$E(J=2^+~T=1)~(MeV)$ & $B(E2)\uparrow$\tablenotemark[4] $(e^2fm^4)$\\ 
\tableline
 2.19 & 4.954\\
 3.40 & 47.170\\
 3.74 & 0.0\\
 7.31 & 0.0\\
 9.16 & 0.0\\
10.92 & 0.0\\
11.92 & 0.0\\
12.12 & 0.1321\\
13.90 & 0.0\\
\end{tabular}
\tablenotetext[1] { Isovector $B(M1)$ from the ground state to excited
$ J=1^+$ states.}
\tablenotetext[2] { Sum of transitions to a four-fold degenerate state
at 12.12 $MeV$.}
\tablenotetext[3] { This state is three-fold degenerate.}
\tablenotetext[4] { Physical $B(E2)$ transitions (with
$e_p=1.0,~e_n=0$) from the ground state to excited $J=2^+$ states.}
\end{table}

\begin{table}
\caption{The $B(E2)$'s (in $e^2fm^4$) and Static Quadrupole
Moments $\rm Q$ (in $efm^2$) of the $2^+_1$ and $2^+_2$ states in
$^{10}Be$ for selected values of $t$.}
\begin{tabular}{ccccc}
 $t$ & $B(E2)_{0^+ \rightarrow 2_1^+}$ & $\rm Q(2^+_1)$ & $B(E2)_{0^+
\rightarrow 2_2^+}$ & $\rm Q(2^+_2)$ \\ 
\tableline
 1.1 & 34.2 & 10.77 & 128.4 & -10.71\\
 1.3 & 43.9 & 11.90 & 172.4 & -11.78\\
 1.5 & 46.4 & 12.12 & 186.6 & -11.98\\
 1.7 & 48.3 & 12.25 & 194.4 & -10.85\\
1.75 & 42.1 & 12.25 &  23.2 & -10.83\\
1.76 &  5.1 & 12.21 & 120.2 & -10.78\\
1.78 & 139.9 & -10.84 & 0 & 12.28\\
 1.8 & 139.9 & -10.84 & 0 & 12.28\\
 2.0 & 139.9 & -10.85 & 0 & 12.34\\
\end{tabular}
\end{table}

\begin{table}
\caption{The Orbital Occupancies in $0^+$ and $2^+$ States of Interest
for Various Values of $t$}
\begin{tabular}{ccccccc}
$t$ & $J^{\pi}$ & $E$ & $0s$ & $0p$ & $1s-0d$ & $0f-1p$\\
\tableline
1.0 & $0^+$ & 0.000 & 3.92 & 5.90 & 0.08 & 0.10\\
    & $0^+$ & 3.746 & 3.92 & 5.88 & 0.09 & 0.11\\
    &       &       &      &      &      &     \\
    & $2^+$ & 2.187 & 3.92 & 5.89 & 0.09 & 0.10\\
    & $2^+$ & 3.400 & 3.93 & 5.90 & 0.08 & 0.09\\
    & $2^+$ & 3.745 & 3.92 & 5.88 & 0.09 & 0.11\\
\tableline

1.3 & $0^+$ & 0.000 & 3.88 & 5.84 & 0.13 & 0.15\\
    & $0^+$ & 4.775 & 3.89 & 5.82 & 0.13 & 0.16\\
    &       &       &      &      &      &     \\
    & $2^+$ & 2.783 & 3.88 & 5.83 & 0.13 & 0.16\\
    & $2^+$ & 4.518 & 3.89 & 5.84 & 0.13 & 0.15\\
    & $2^+$ & 4.775 & 3.89 & 5.82 & 0.13 & 0.16\\
\tableline

1.5 & $0^+$ & 0.000 & 3.86 & 5.81 & 0.16 & 0.18\\
    & $0^+$ & 2.890 & 4 & 4 & 2 & 0\\
    &       &       &      &      &      &     \\
    & $2^+$ & 3.174 & 3.86 & 5.80 & 0.16 & 0.18\\
    & $2^+$ & 5.191 & 3.86 & 5.79 & 0.16 & 0.19\\
    & $2^+$ & 5.447 & 3.86 & 5.79 & 0.16 & 0.19\\
\tableline

1.7 & $0^+$ & 0.000 & 3.84 & 5.78 & 0.18 & 0.20\\
    & $0^+$ & 0.585 & 4 & 4 & 2 & 0\\
    &       &       &      &      &      &     \\
    & $2^+$ & 3.562 & 3.84 & 5.77 & 0.18 & 0.21\\
    & $2^+$ & 4.333 & 4 & 4 & 2 & 0\\
    & $2^+$ & 5.805 & 3.84 & 5.75 & 0.19 & 0.22\\
\tableline

1.9 & $0^+$ & 0.000 & 4 & 4 & 2 & 0\\
    & $0^+$ & 1.564 & 3.82 & 5.76 & 0.20 & 00.22\\
    &       &       &      &      &      &     \\
    & $2^+$ & 4.190 & 4 & 4 & 2 & 0\\
    & $2^+$ & 5.512 & 3.82 & 5.74 & 0.20 & 0.23\\
    & $2^+$ & 7.934 & 3.82 & 5.72 & 0.21 & 0.25\\
\end{tabular}
\end{table}

\end{document}